\documentclass[aps,floats,nofootinbib,amssymb,preprint,tighten,superscriptaddress]{revtex4}
\usepackage{epsf,epsfig}
\usepackage{subfigure}
\usepackage{graphicx}
\usepackage{color}

\newcommand{\be}{\begin{equation}}
\newcommand{\ee}{\end{equation}}
\newcommand{\bea}{\begin{eqnarray}}
\newcommand{\eea}{\end{eqnarray}}

\renewcommand{\check}{({\bf This statement needs to be checked!!})}

\newcommand{\ba}{\begin{array}}
\newcommand{\ea}{\end{array}}
\newcommand{\bi}{\begin{itemize}}
\newcommand{\ei}{\end{itemize}}
\newcommand{\ben}{\begin{enumerate}}
\newcommand{\een}{\end{enumerate}}

\preprint{
\hbox to \hsize{
\hfill$\vcenter{\hbox{\bf MAD-PH-09-1548}
         \hbox{\bf NUHEP/09-012}
         \hbox{\bf ANL-HEP-PR-09-106}
         \hbox{November 2009}}$}
}

\begin{document}
\title{\vspace*{.75in}
Single top and Higgs associated production at the LHC}
\author{Vernon Barger}
\affiliation{Department of Physics, University of Wisconsin, Madison, WI 53706}
\author{Mathew McCaskey}
\affiliation{Department of Physics, University of Wisconsin, Madison, WI 53706}
\author{Gabe Shaughnessy}
\affiliation{Northwestern University, Department of Physics and Astronomy, Evanston, IL 60208 USA}
\affiliation{HEP Division, Argonne National Lab, Argonne IL 60439 USA}

\thispagestyle{empty}
\begin{abstract}
We study the production of a Standard Model (SM) Higgs boson in association with a single top quark and either a light jet or $W$-boson at the LHC with a center of mass energy of $14$ TeV.  Due to the destructive interference of the contributing SM diagrams, the value of the top Yukawa coupling and the sign of the $WWh$ coupling may be probed for Higgs masses above $150$ GeV where $WW$ and $ZZ$ are the dominant Higgs decays.  We consider Higgs masses of $m_{h}=120$, $150$, $180$, and $200$ GeV and devise experimental cuts to extract the signal from SM backgrounds and measure the top Yukawa coupling.
\end{abstract}
\maketitle

\newpage

\section{Introduction}

The Higgs boson associated with spontaneous electroweak symmetry breaking (EWSB) is one of the most important anticipated discoveries in modern particle physics.  To this end, several Higgs boson production and decay channels have been extensively studied.   Searches at the Fermilab Tevatron have recently excluded a Standard Model (SM) Higgs boson in the mass range of $160$-$170$ with $95\%$ confidence GeV~\cite{:2009pt}, wherein $H\to WW^{(*)}$ is the dominant decay mode.   Simulations for upcoming searches by the ATLAS and CMS detectors at the LHC~\cite{RichterWas:2009wx,Bayatian:2006zz,Ball:2007zza} at $14$ TeV indicate that a $5$ sigma Higgs discovery is achievable with $10$ fb$^{-1}$ luminosity over the full SM Higgs boson mass range of interest, $110$ - $800$ GeV~\cite{Mangano:2009ph}.  

The upper bound of this mass range comes from the requirement that the electroweak (EW) theory is consistent up to a certain energy scale $\Lambda$.  By analyzing the running of the quartic coupling in the Higgs potential up to $\Lambda$, an upper bound can be put on the coupling and hence the Higgs mass itself~\cite{Hambye:1996wb,Quigg:2009vq}.  If electroweak theory is consistent up to the Plank scale this upper bound is $\lesssim 200$ GeV~\cite{Tobe:2002zj}.  Lowering the energy scale of new physics to $\Lambda = 1$ TeV, accessable at the LHC, the upper bound on the SM Higgs mass increases to $\lesssim 800$ GeV.  The lower bound on the Higgs mass arises from the stability requirement of the Higgs potential.  Radiative corrections associated with top-quark loops can destabilize the minimum of the Higgs potential if the Higgs mass is $<50$ GeV at $\Lambda$ = $1$ TeV and $<100$ GeV at the Planck scale~\cite{Casas:1996aq,Einhorn:2007rv,Quigg:2009vq}.  A SM Higgs with mass less than $114$ GeV is excluded by LEP 2 searches~\cite{Barate:2003sz}.

Beyond Higgs discovery, measurements of the Higgs couplings to other SM particles will test the fundamental properties of the Higgs~\cite{Djouadi:2000gu,Zeppenfeld:2000td,Belyaev:2002ua,Djouadi:2007ik,Barger:2009me}.  Its couplings to the $W$ and $Z$ bosons are definitive tests of the SM, since both the $W$, $Z$ masses and the $WWh$, $ZZh$ couplings are determined by the vacuum expectation value (vev) of the neutral physical Higgs state.  The relation of the $W$-boson mass to the vev including the radiative corrections from the top quark and Higgs boson loops is~\cite{Sirlin:1980nh,Hollik:1988ii,Dawson:2008jk}\footnote{This approximation is only valid for $m_{h}\gg M_{Z}$.}

\begin{equation}
M_{W}^{2}=\frac{\pi\alpha}{\sqrt{2}G_{F}s_{w}^{2}}\left[1.07-\frac{G_{F}}{\sqrt{2}}\frac{3}{8\pi^{2}}\left(\frac{c_{w}^{2}}{s_{w}^{2}}\right)m_{t}^{2}+\frac{\alpha}{\pi s_{w}^{2}}\frac{11}{48}\left[\log\left(\frac{M_{h}^{2}}{M_{Z}^{2}}\right)-\frac{5}{6}\right]\right]
\end{equation}
where $c_{w}$ and $s_{w}$ are the cosine and sine of the weak mixing angle, respectively.

Measurements of the Yukawa couplings of the Higgs boson to the top and bottom quarks will show whether the third generation quark masses are also generated by the SM Higgs.  More generally, measurements of the weak boson and fermion couplings can differentiate Higgs mechanisms that involve more than one Higgs doublet~\cite{Barger:2009me}.

The principal SM Higgs production mechanisms are $Z$ and $W$ Higgstrahlung~\cite{Glashow:1978ab,Finjord:1979fc,Eichten:1984eu}, Weak Boson Fusion~\cite{Jones:1979bq,Cahn:1983ip,Asai:2004ws,Cranmer:2004uz}, gluon fusion~\cite{Georgi:1977gs,Duhrssen:2004uu,Duhrssen:2004cv}, and production in association with a top quark pair~\cite{Raitio:1978pt,Kunszt:1984ri,Bagdasaryan:1986fj,Ng:1983jm,Maltoni:2002jr,Su:2008bj,Lafaye:2009vr,Lipatov:2009qx}.   The gluon fusion process occurs at loop level and as such that prediction is dependent on the contributions of the virtual SM particles in the loops, dominated by the top-quark loop.   The SM Higgs boson decay branching fractions depend sensitively on the Higgs mass, with the $WW^{(*)}$ and $ZZ^{(*)}$ decay modes dominating above the $WW$ threshold and $b\bar{b}$ dominating at lower Higgs masses.

In principle, the top quark Yukawa coupling may be probed through Higgs production with an associated top, anti-top quark pair~\cite{Raitio:1978pt,Kunszt:1984ri,Bagdasaryan:1986fj,Ng:1983jm,Maltoni:2002jr,Su:2008bj,Lafaye:2009vr,Lipatov:2009qx,Plehn:2009rk} and Higgs production via gluon fusion~\cite{Georgi:1977gs,Duhrssen:2004uu,Duhrssen:2004cv}. These processes require the high energy and high luminosity of the LHC.  Early studies done to isolate the Higgs signal in the $t\bar{t}h$ production process found optimistic conclusions but more refined simulations now indicate that it will be very difficult to isolate this signal from SM backgrounds for a Higgs mass in the range of $120$-$200$ GeV~\cite{RichterWas:2009wx}. However, a more optimistic assessment has recently been reached in Ref.~\cite{Plehn:2009rk}.

Our interest in this paper is the potential LHC measurement of the top Yukawa coupling through Higgs production in association with a single top quark.  Previous simulations of this process and its SM backgrounds~\cite{Maltoni:2001hu,DiazCruz:1991cs,Stirling:1992fx,Ballestrero:1992bk,Bordes:1992jy,Tait:2000sh,Campbell:2009ss}  have focused on SM Higgs masses for which the $b\bar{b}$ decay mode dominates.   We revisit the $H\to b\bar{b}$ channel and reproduce the results of previous simulations that this signal is buried by backgrounds.  Thereafter we focus on the equally interesting case where the Higgs decays dominantly to the $WW^{(*)}$ and $ZZ^{(*)}$ final states.

The single top-Higgs channel provides a unique way to test the SM prediction of the sign of the $WWh$ coupling due to the interference of two contributing Feynman diagrams~\cite{Tait:2000sh}: see Section~\ref{sec:singletopintro}.  We discuss our collider simulations and acceptance cuts to extract the $H\to WW^{(*)},ZZ^{(*)}$ signals from SM backgrounds in Sections~\ref{sec:collsim} and~\ref{sec:collstudy} respectively.  Finally, in Section~\ref{sec:beyondtheSM} we quantitatively evaluate the interference between the contributing diagrams to test the sign of the $WWh$ coupling.

We base our study on the center of mass design energy of $14$ TeV and an integrated luminosity of $300$ fb$^{-1}$ for each of the two detectors (ATLAS and CMS).  The super-LHC would deliver ten times the luminosity of the LHC~\cite{Mangano:2009ph} and accordingly should increase the sensitivity to the process of interest here.

\section{Single Top Quark and Higgs Associated Production}
\label{sec:singletopintro}

\begin{figure}[htpb]
\begin{center}
\subfigure[]{\includegraphics[angle=0,width=0.18\textwidth]{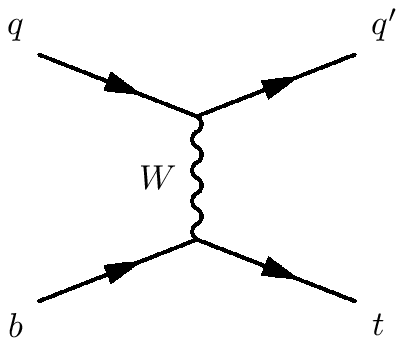}}~~~
\subfigure[]{\includegraphics[angle=0,width=0.20\textwidth]{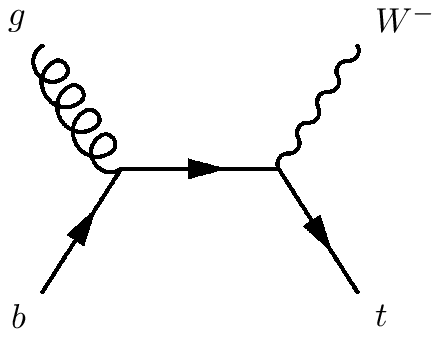}}~~~
\subfigure[]{\includegraphics[angle=0,width=0.20\textwidth]{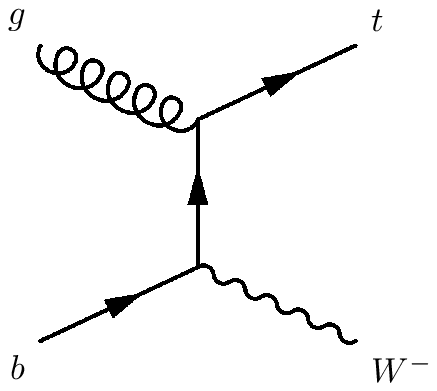}}~~~
\subfigure[]{\includegraphics[angle=0,width=0.20\textwidth]{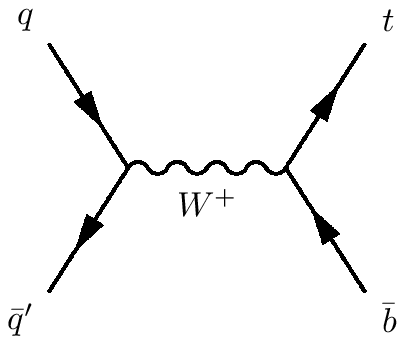}}
\caption{Feynman diagrams for single top production at the LHC with various associated particles. (a) The $t$-channel $W$ exchange with an associated light jet. (b-c) The $s$ and $t$ channel diagrams with an associated $W$-boson. (d) The $s$-channel $W$ exchange with an associated $b$ jet. Similar diagrams exist for anti-top quarks in the final state.}
\label{fig:SingletopFD}
\end{center}
\end{figure}

Single top production has been studied extensively both in the SM and models beyond the SM~\cite{Cao:2007ea,Alan:2007ui,LopezVal:2008zu,AguilarSaavedra:2008gt,Wang:2008iw,Coimbra:2008qp,Kribs:2009zy,Plehn:2009it,Aliev:2009jm}. In the SM a single top quark can be produced in association with either a light quark, $W$-boson, or bottom quark.  The Feynman diagrams for these three processes are given in Fig.~\ref{fig:SingletopFD}.

The single top production process is of electroweak strength. Nevertheless, at the LHC the cross section is within a factor of $5$ of $t\bar{t}$ production via the strong process $gg\to t\bar{t}$.  We concentrate on the single top and single anti-top channels with an associated light quark and associated $W$-boson, since those processes have the largest cross sections: see Table~\ref{tab:topcross}. In addition, we note that the large $t\bar{t}$ background process, largely from $gg\to t\bar{t}$ at LHC energies, can mimic the $tb$ process with a failed lepton tag.

\begin{table}[htpb]
\begin{center}
\begin{tabular}{|cc|cc|}
\hline
$\sigma(pp\to tj)$ & $120$ pb & $\sigma(pp\to \bar{t}j)$ & $67$ pb \\
$\sigma(pp\to tW^{-})$ & $31$ pb & $\sigma(pp\to \bar{t}W^{+})$ & $31$ pb\\
$\sigma(pp\to tb)$ & $4.0$ pb & $\sigma(pp\to \bar{t}\bar{b})$ & $2.4$ pb\\
\hline
$\sigma(pp\to t\bar{t})$ & $582$ pb & & \\
\hline
\end{tabular}
\caption{Comparison of top pair and single top production cross sections at $14$ TeV center of mass energy.  All of the cross sections here and in the rest of our paper were calculated using the MadGraph~\cite{Alwall:2007st} software package with the CTEQ6L parton distribution functions~\cite{Pumplin:2005rh}. All cross sections were calculated to have a statistical error less than $1\%$. }
\label{tab:topcross}
\end{center}
\end{table}

Several calculations have been made of next-to-leading order (NLO) and next-to-next-to-leading order (NNLO) corrections to single top production at the LHC~\cite{Campbell:2009gj,Bonciani:2009dh,Heim:2009ku,Kidonakis:2007ej,Bordes:1994ki,Stelzer:1997ns,Stelzer:1998ni,Tait:1999cf}.  These corrections can increase the production cross section by as much as $50\%$ in the case of the $tW$ channel; the corrections are more modest for the $tq'$ and $t\bar{b}$ channels.  The higher order QCD corrections to single top-Higgs associated production will be similar.  For this study we conservatively concentrate on the contributions from leading order diagrams without QCD K-corrections.

\begin{figure}[b]
\begin{center}
\subfigure[]{\includegraphics[angle=0,width=0.18\textwidth]{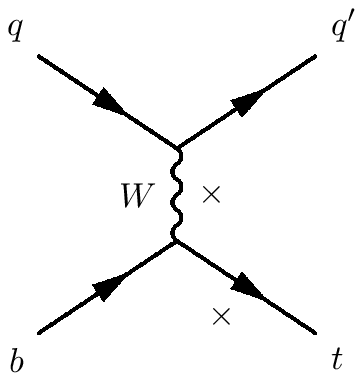}}~~~
\subfigure[]{\includegraphics[angle=0,width=0.20\textwidth]{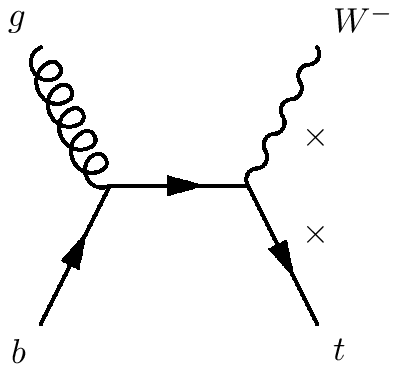}}~~~
\subfigure[]{\includegraphics[angle=0,width=0.20\textwidth]{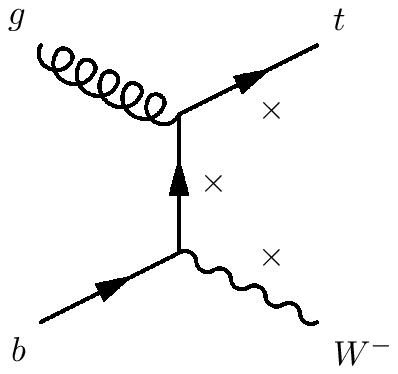}}~~~
\subfigure[]{\includegraphics[angle=0,width=0.20\textwidth]{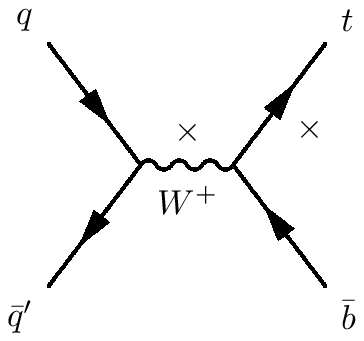}}
\caption{Feynman diagrams contributing to the Higgs production with an associated single top. The diagrams are the same as the single top production but with a Higgs boson radiated from the $W$-boson or the top quark.  The locations where the Higgs can be radiated is denoted by $\times$ with only one radiated Higgs implied.}
\end{center}
\label{fig:SingletopHiggsFD}
\end{figure}

To produce a SM Higgs in association with a single top quark, a Higgs boson is radiated from each of the massive particles in the Feynman diagrams associated with single top production.  Higgs radiation from a $b$ quark (or any other light quark) is suppressed by the small Yukawa coupling and the intermediate quark being far off-shell.  Therefore, Higgs radiation from the $W$-boson and the top quark give the dominating contributions to the production cross section.  The extended Feynman diagrams, including the sites where the Higgs is radiated, are given in Fig.~\ref{fig:SingletopHiggsFD}, with $\times$ marking possible Higgs emissions.

We consider Higgs masses of $120$, $150$, $180$, and $200$ GeV in our analysis.  The low Higgs mass of $120$ GeV, with $h\to b\bar{b}$ decay, has been studied previously in~\cite{Maltoni:2001hu} and was shown to have insurmountable background contamination.  The other Higgs mass cases have large branching fractions to weak bosons which potentially makes it easier to extract the signal from the background when one of the weak bosons decay leptonically.  We calculate the production cross sections for the aforementioned Higgs masses and obtain the results given in Table~\ref{tab:higgssingletopcross}.  The produced top (anti-top) quarks in association with a Higgs boson that is radiated from the $t$($\bar{t}$) prefferentially have a right-handed (left-handed) polarization, due to the $V-A$ weak coupling and the chirality flip from the Higgs emission.  The leptons from $t_{L}$ and $t_{R}$ semileptonic decays have different distributions, but the fact that both tops and anti-tops are produced makes it difficult to exploit the differences.

\begin{table} [htpb]
\begin{center}
\begin{tabular}{|c|cc|cc|cc|c|}
\hline
 & $tjh$ & $\bar{t}jh$ & $tW^{-}h$ & $\bar{t}W^{+}h$ & $tbh$ & $\bar{t}\bar{b}h$ & $t\bar{t}h$ \\
\hline
$m_{h}=120$ GeV & $45$ & $23$ & $9.0$ & $9.0$ & $1.6$ & $0.8$ & $440$ \\
$m_{h}=150$ GeV & $33$ & $19$ & $5.0$ & $5.0$ & $1.0$ & $0.5$ & $240$ \\
$m_{h}=180$ GeV & $31$ & $16$ & $3.0$ & $3.0$ & $0.6$ & $0.3$ & $140$ \\
$m_{h}=200$ GeV & $29$ & $15$ & $2.4$ & $2.4$ & $0.5$ & $0.2$ & $100$ \\
\hline
\end{tabular}
\caption{The cross sections (in fb) at $14$ TeV of the SM Higgs production with an associated single top and the three possible final state particles for various Higgs mass choices.  The cross section for the Higgs production with an associated top quark pair is included in the last column for comparison.}
\label{tab:higgssingletopcross}
\end{center}
\end{table}

We calculate the branching fractions of the Higgs to SM paticles using HDECAY~\cite{Djouadi:1997yw} for the different Higgs masses.  These results are given in Table~\ref{tab:bfs}.

\begin{table}[htpb]
\begin{center}
\begin{tabular}{|c|ccc|}
\hline
$m_{h}$ (GeV) & BF($h\to ZZ^{(*)}$) & BF($h\to WW^{(*)}$) & BF($h\to b\bar{b}$) \\
\hline
$120$ & $0.015$ & $0.138$ & $0.673$ \\
$150$ & $0.082$ & $0.693$ & $0.168$ \\
$180$ & $0.059$ & $0.933$ & $0.005$ \\
$200$ & $0.253$ & $0.743$ & $0.003$ \\
\hline
\end{tabular}
\caption{Branching fractions of the Higgs to $ZZ^{(*)}$, $WW^{(*)}$, and $b\bar{b}$ pairs for the Higgs masses we study.}
\label{tab:bfs}
\end{center}
\end{table}

\section{Collider Simulations}
\label{sec:collsim}

The events used in our analysis were generated at tree level with the MadGraph~\cite{Alwall:2007st} and Alpgen~\cite{Mangano:2002ea} packages using the CTEQ6L parton distribution functions~\cite{Pumplin:2005rh}.  We use a renormalization and factorization scale consistent with Refs.~\cite{Dawson:2005vi,Kao:2009jv} given by

\begin{equation}
Q=\frac{1}{2}\sum_{i=t,h,W,Z}M_{i}.
\end{equation}

To simulate the unweighted events at the parton level we apply energy and momentum smearing to the final state particles according to the usual prescription

\begin{equation}
\frac{\delta E}{E}=\frac{a}{\sqrt{E}}\oplus b,
\label{eq:smearing}
\end{equation}

\noindent where $a=0.5, 0.1$ and $b=0.03, 0.007$ for jets and leptons respectively.  We require the final state particles to have $p_{T}>p_{T}^{min}$ and $|\eta|<\eta_{max}$, where $p_{T}^{min}$ and $\eta_{max}$ are defined as

\begin{eqnarray}
\begin{tabular}{l}
$p_{T}^{\ell=e,\mu}>20$~GeV, \qquad $|\eta_{e}|<2.4$, \qquad $|\eta_{\mu}|<2.1$, \\
$p_{T}^{\tau}>15$~GeV, \qquad $|\eta_{\tau}|<2.5$, \\
$p_{T}^{j,b}>20$~GeV, \qquad $|\eta_{j}|< 4.5$, \qquad $|\eta_{b}|<2.0$.
\end{tabular}
\end{eqnarray}

\noindent Jets are defined at the parton level.  We also require final state particles to be isolated in the detector.  Our isolation requirement is a minimum $\Delta R$ between all final leptons and jets as defined by

\begin{equation}
\Delta R = \sqrt{\Delta\eta^{2}+\Delta\phi^{2}}.
\end{equation}

\noindent where $\Delta\eta$ is the rapidity gap and $\Delta\phi$ is the azimuthal angle gap between the particle pair.  We impose the following criteria for $\Delta R$ separations:

\begin{equation}
\Delta R(jj, j\ell, \ell\ell) > 0.4.
\end{equation}

Along with these acceptance and isolation criteria, we impose a tagging efficiency of $0.6$, $0.9$, and $0.4$ for final state $b$ jets, leptons (electrons and muons), and taus respectively.  We also include $b$ and $\tau$ mis-tagging rates as follows:

\begin{eqnarray}
\varepsilon_{c\to b}&=&0.1~\text{for}~p_{T}>50~\text{GeV} \\
\varepsilon_{j\to b}&=&\left\{ \begin{array}{cl}
 \frac{1}{50} & \mbox{ for $p_{T}>250$ GeV} \\
 \frac{1}{150}\left(\frac{2(p_{T}-100)}{150}+1\right) & \mbox{ for $100$ GeV $<p_{T}<250$ GeV} \\
 \frac{1}{150} & \mbox{ for $p_{T}<100$ GeV} 
 \end{array} \right. \\
\varepsilon_{q\to\tau}&=&\left\{ \begin{array}{cl}
 \frac{1}{30} & \mbox{ for $15$ GeV $<p_{T}<30$ GeV} \\
 \frac{1}{100} & \mbox{ for $30$ GeV $<p_{T}$} 
 \end{array} \right.
\end{eqnarray}

\noindent These tagging and mis-tagging efficiencies are consistent with the values in the latest ATLAS TDR~\cite{RichterWas:2009wx}.

\section{Collider Study}
\label{sec:collstudy}

\subsection{Backgrounds}

The main Higgs decay channels are weak boson or bottom quarks pairs (See Table~\ref{tab:bfs}).  The further hadronic decay of the weak bosons leads to at most four jets from the Higgs, neglecting jets from final state radiation.  With the Higgs produced in conjunction with a top quark and either a light jet or a $W$-boson, the hadronic decays of these particles lead up to a total of eight or nine jets respectively in the final state.  The QCD background with the largest cross section is a $t\bar{t}$ pair produced with up to two extra jets.  A more complete set of backgrounds include electroweak processes that give similar number of final state leptons and jets.  We include $t\bar{t}V$, $t\bar{t}Vj$, $tVV$, $VVV$, $VVVj$, $VVVV$, and $VVVVj$ in our list of background events, where $V=W,Z$.  A summary of the cross sections of all background processes considered is given in Table~\ref{tab:backgrounds}.  The QCD and EW background processes are given along with their cross sections and maximum numbers of jets that can appear in the final state. 

\begin{table}[htpb]
\begin{center}
\begin{tabular}{|c|c|c|c|c|c|c|c|c|}
\hline
Process & $t\bar{t}$ & $t\bar{t}j$ & $t\bar{t}jj$ & $t\bar{t}Vj$ & $t\bar{t}V$ & $3Vj,4Vj$ & $3V,4V$ & $tVV$ \\
\hline
$\sigma$(pb) & $530$ & $440$ & $300$ & $1.2$ & $1.1$ & $.37$ & $.27$ & $.12$ \\
max($N_{jets}$) & $6$ & $7$ & $8$ & $9$ & $8$ & $7-9$ & $6-8$ & $7$ \\ 
\hline
\end{tabular}
\caption{Cross sections for SM background processes included in our analysis for the same parton distribution functions and acceptance cuts as our signal calculations.  These cross sections and maximum number of jets enable us to decide which background channels will be important for each of the signal processes.}
\label{tab:backgrounds}
\end{center}
\end{table}

\subsection{$h\to b\bar{b}$ Channel}

The $tjh$ signal with a Higgs mass of $120$ GeV has been studied in Ref~\cite{Maltoni:2001hu} for the SM and in Ref~\cite{Yang:2009ff} for the little Higgs model.  For $h\to b\bar{b}$ at $m_{h}=120$ GeV~\footnote{We cannot exploit the loop induced decay $H\to\gamma\gamma$ due to its low branching fraction ($\mathcal{O}(10^{-3})$).}, the primary backgrounds are $t\bar{t}$ and $t\bar{t}j$ for the $tjh$ signal and $t\bar{t}j$ and $t\bar{t}jj$ for the $tWh$ signal.

To reduce the QCD backgrounds we require a tagged lepton from the semileptonic decay of the associated top quark.  Because the $tjh$ process is a $t$-channel $W$ exchange the associated light jet will nominally have high rapidity and high $p_{T}$.  We therefore make a cut requiring a light quark jet (i.e. not a $b$-tagged jet) to have an absolute pseudorapidity greater than $2$ and a $p_{T}$ greater than $50$ GeV.  We reconstruct the Higgs mass from two of the three tagged $b$ jets and require the invariant mass of these two $b$-jets to be within a $40$ GeV window centered around the Higgs mass to allow for resolution of smeared jets.  After imposing these acceptance cuts we find the signal is still overwhelmed by QCD backgrounds as previously found in Ref.~\cite{Maltoni:2001hu}.

In the $tWh$ search we can obtain the required tagged lepton from the associated $W\to \ell\nu$ decay.  This allows us to reconstruct the top quark using a third tagged b-jet and two lighter jets from the $W\to q\bar{q}$ decay.  We take a mass window of $40$ GeV for the reconstruction of both the Higgs (from two $b$-jets) and the top (from a $b$-jet and two light jets) and a window of $20$ GeV for the reconstruction of the $W$-boson from the top decay.  Again we find that the signal is overwhelmed by QCD backgrounds after applying all of our cuts with an integrated luminosity of $600$ fb$^{-1}$.

\subsection{$h\to WW^{(*)}$}

Higgs masses with $m_{h}\gtrsim 2M_{W}$ have large branching fractions to two W bosons (see Table~\ref{tab:bfs}).  The subsequent $W$-boson decays may increase the number of final state particles and the increase in the number of intermediate particles to reconstruct helps make the backgrounds managable.  In events for which both $W$-bosons decay leptonically we could use cuts to exploit the spin-correlations of the leptons;  because both $W$-bosons originate from a spin-$0$ Higgs and the $V-A$ coupling of the $W$-bosons decays make it likely that the two leptons will be detected close together~\cite{Abazov:2005un}.  Unfortunately, the square of the leptonic $W$ branching fraction renders the spin-correlated lepton signal too small to be useful with our small signal cross sections.

For the $tjh$ signal we require a single tagged lepton to reduce the QCD backgrounds.  We consider the case that the lepton originates from the top quark decay to allow full reconstruction of the Higgs.  We require a light jet to have a high rapidity and high $p_{T}$ because of the nature of the $t$ channel $W$ exchange of the signal process.  We can reconstruct two $W$-bosons from the Higgs decay (in the case of the $150$ GeV Higgs only the real $W$-boson can be reconstructed) and the Higgs mass from the two reconstruced $W$-bosons.  Still, we find that for all the Higgs choices the $tjh$ signal is overwhelmed by backgrounds.

\begin{figure}[htpb]
\begin{center}
\includegraphics[angle=0,width=0.6\textwidth]{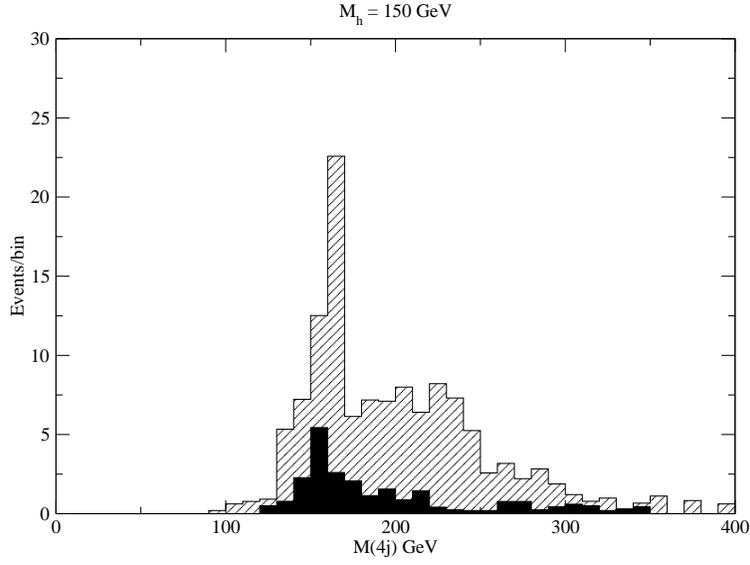}
\caption{Reconstruction of a $150$ GeV Higgs mass via the invariant mass of $4$ jets using $600$ fb$^{-1}$ of data.  A peak at $150$ GeV is visible in the signal (black) within $20$ GeV of the Higgs mass.  The background (hatched) peak is a combinatoric result of the cuts.}
\label{fig:4jhiggsrecon}
\end{center}
\end{figure}

For the $tWh$ signal we next consider the case where the tagged lepton comes from the decay of the associated $W$-boson.  With both the top quark and Higgs decaying to jets we can reconstruct the Higgs boson mass from $4$ light jets.  In this reconstruction we require that the invariant mass of the four jets must be within $20$ GeV of the Higgs mass.  To justify this mass window we take a sample of signal events and plot the invariant mass of $4$ light jets under the condition that the remaining $3$ jets (one of which is a $b$-jet) reconstruct the top quark.  Fig.~\ref{fig:4jhiggsrecon} shows a noticable signal peak centered around $150$ GeV within $20$ GeV of the Higgs mass.

\begin{figure}[htpb]
\begin{center}
\includegraphics[angle=0,width=0.4\textwidth]{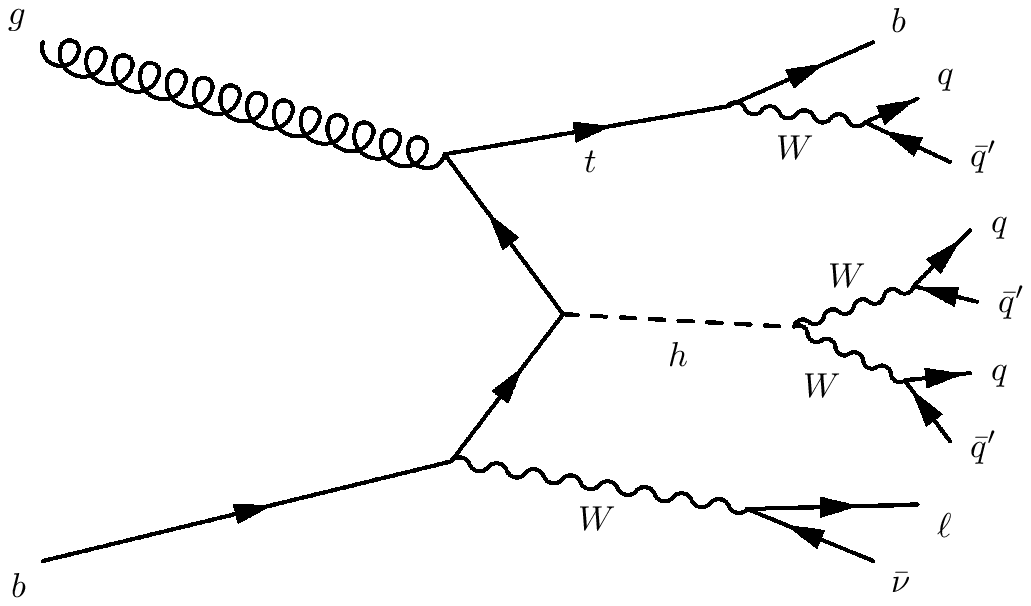}~~~
\includegraphics[angle=0,width=0.4\textwidth]{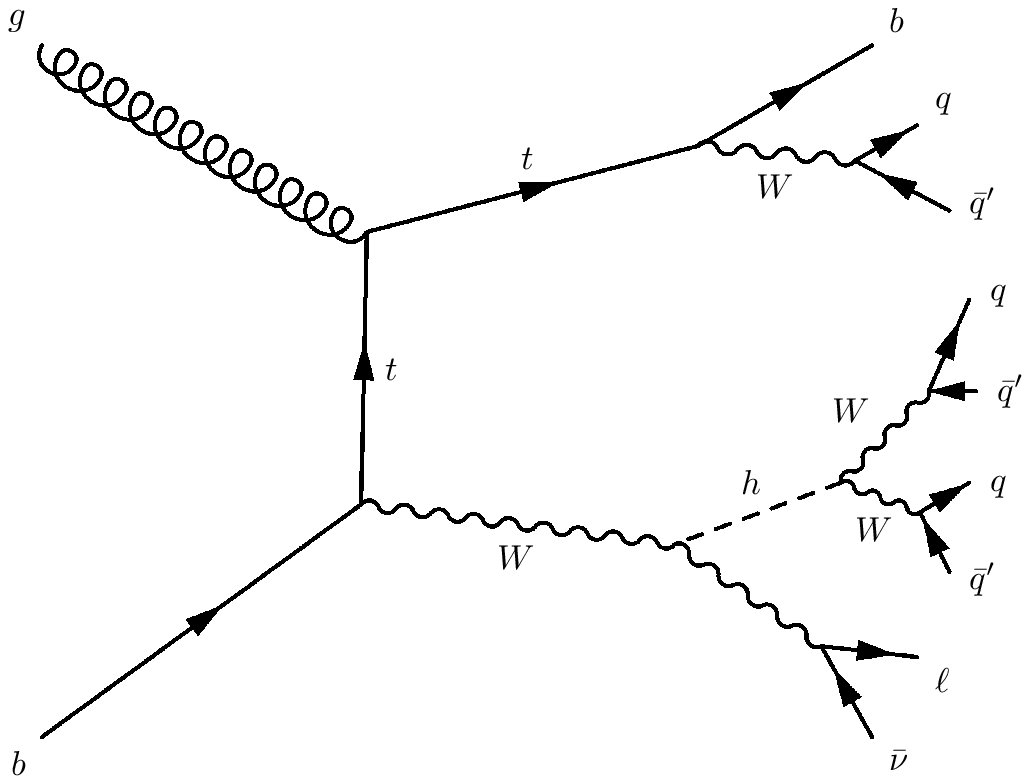}
\caption{Extended Feynman diagrams for the $tWh$ signal where the $h\to WW$ decay channel is shown.}
\label{fig:tWhFD}
\end{center}
\end{figure}

Next we reconstruct the top quark from the $3$ remaining jets of the signal by requiring the invariant mass of the jets to be within $20$ GeV of the top mass.  To further reduce backgrounds we reconstruct the $W$-boson from the top quark decay and the real $W$-bosons from the Higgs decay (In the $m_{h}= 150$ GeV case only one of the $W$ is reconstructable.).  The Feynman Diagrams for this signal are shown in Fig.~\ref{fig:tWhFD}.  

With these cuts we obtain the event rates given in Table~\ref{tab:htoWW-tWh}.  Because the $tWh$ signal yields one more final state jet than the $tjh$ signal, many of the QCD backgrounds that overwhelmed the $tjh$ signal do not pass our set of cuts.  The dominant background for this process is $t\bar{t}Vj$.  With $600$ fb$^{-1}$ of integrated luminosity we obtain a signal statistical significance $S/\sqrt{(S+B)}$ of $1.6$, $1.2$, and $0.7$ for $150$, $180$, and $200$ GeV Higgs masses respectively.  The smaller statistical significance for the signal of the $200$ GeV Higgs mass can be attributed to the smaller production cross section and a larger number of background events passing the Higgs mass reconstruction cut.

\begin{table}[tpb]
\begin{center}
\begin{tabular}{|c|c|c|c|c|c|c|c|c|c|}
\hline
Signal Event Rate at $600$ fb$^{-1}$ & \multicolumn{3}{c|}{$m_{h}=150$ GeV} & \multicolumn{3}{c|}{$m_{h}=180$ GeV} & \multicolumn{3}{c|}{$m_{h}=200$ GeV} \\
for $7j+1\ell$: &  \multicolumn{3}{c|}{270 events} & \multicolumn{3}{c|}{230 events} & \multicolumn{3}{c|}{140 events} \\
\hline
Cut & $S$ & $B$ & $\frac{S}{\sqrt{S+B}}$ & $S$ & $B$ & $\frac{S}{\sqrt{S+B}}$ & $S$ & $B$ & $\frac{S}{\sqrt{S+B}}$ \\
\hline
tagging: $7$ j, $1\ell$, $1$ b & $67$ & $20100$ & $0.5\pm0.1$ & $64$ & $20100$ & $0.4\pm0.1$ & $50$ & $20100$ & $0.4\pm0.1$ \\
$|M_{4j}-m_{h}|<20$ GeV & $44$ & $2000$ & $1.0\pm0.2$ & $45$ & $5200$ & $0.6\pm0.1$ & $41$ & $7600$ & $0.5\pm0.1$ \\
$|M_{bjj}-m_{t}|<20$ GeV & $24$ & $390$ & $1.2\pm0.3$ & $26$ & $1100$ & $0.8\pm0.2$ & $23$ & $1800$ & $0.5\pm0.1$ \\
$|M_{2j(top)}-M_{W}|<10$ GeV & $21$ & $160$ & $1.6\pm0.3$ & $25$ & $520$ & $1.1\pm0.2$ & $17$ & $890$ & $0.6\pm0.1$ \\
$|M_{2j(Higgs)}-M_{W}|<20$ GeV & $20$ & $150$ & $1.5\pm0.3$ & $23$ & $320$ & $1.2\pm0.3$ & $16$ & $580$ & $0.7\pm0.2$ \\
\hline
\end{tabular}
\caption{Cuts used to extract the $tWh$ signal with $h\to WW^{(*)} \to$ jets for Higgs masses of $150$, $180$, and $200$ GeV.  The event rates include the appropriate branching fractions for a final state of $7$ jets and $1$ lepton. With each sequential cut the number of signal and background events that pass each sequential cut are given along with the resulting statistical significance of the signal. The statistical significance uncertainty is given by the propagated poisson uncertainties in the signal ($S$) and background ($B$) events.}
\label{tab:htoWW-tWh}
\end{center}
\end{table}

\subsection{$h\to ZZ^{(*)}$}

\begin{figure}[tpb]
\begin{center}
\includegraphics[angle=0,width=0.6\textwidth]{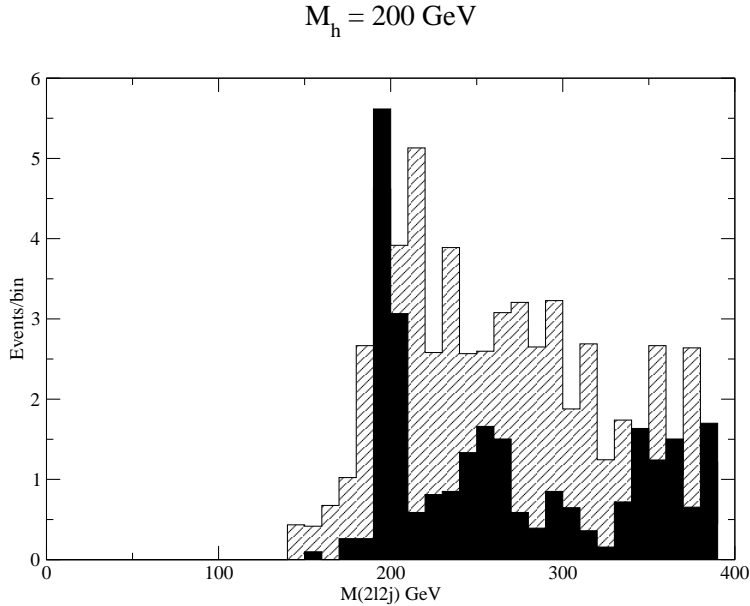}
\caption{Reconstruction of a $200$ GeV Higgs mass from the invariant mass of $2$ leptons and $2$ jets using $600$ fb$^{-1}$ of data.  Signal events are denoted by black histograms and background events by the hatched histogram.  With each event plotted the lepton and jet pairs both reconstruct a $Z$ boson and a top quark is reconstructed from $2$ other light jets and a $b$-tagged jet.  A peak at $200$ GeV is visible within $10$ GeV of the Higgs mass.}
\label{fig:2j2l-higgsrecon}
\end{center}
\end{figure}

Though the branching fraction of $h\to ZZ^{(*)}$ is smaller than the branching fraction of $h\to WW^{(*)}$ (Table~\ref{tab:bfs}), the $ZZ^{(*)}$ signal is much easier to separate from SM backgrounds.  If one of the $Z$ bosons from the Higgs decays leptonically then we can reconstruct that $Z$ very precisely because the lepton energy smearing is small (Eq.~\ref{eq:smearing}).  This precise reconstruction of the $Z$ boson greatly reduces the QCD backgrounds ($t\bar{t}$, $t\bar{t}j$, $t\bar{t}jj$, etc.) as it effectively requires a $Z$ boson to be on-shell.  For the second $Z$ we require two jets to have an invariant mass within a $20$ GeV window centered around $M_{Z}$, except for a $150$ GeV Higgs where we can only reconstruct one of the $Z$ bosons as the other will be off-shell.  The two pairs of leptons and jets reconstruct the Higgs mass by requiring the invariant mass of the $2$ jets and $2$ leptons to be within $10$ GeV of the Higgs mass.  Fig.~\ref{fig:2j2l-higgsrecon} illustrates why this mass window was chosen as this mass distribution shows a sharp peak in the invariant mass of the $2$ jets and $2$ leptons centered around the Higgs mass.

\begin{figure}[htpb]
\begin{center}
\includegraphics[angle=0,width=0.4\textwidth]{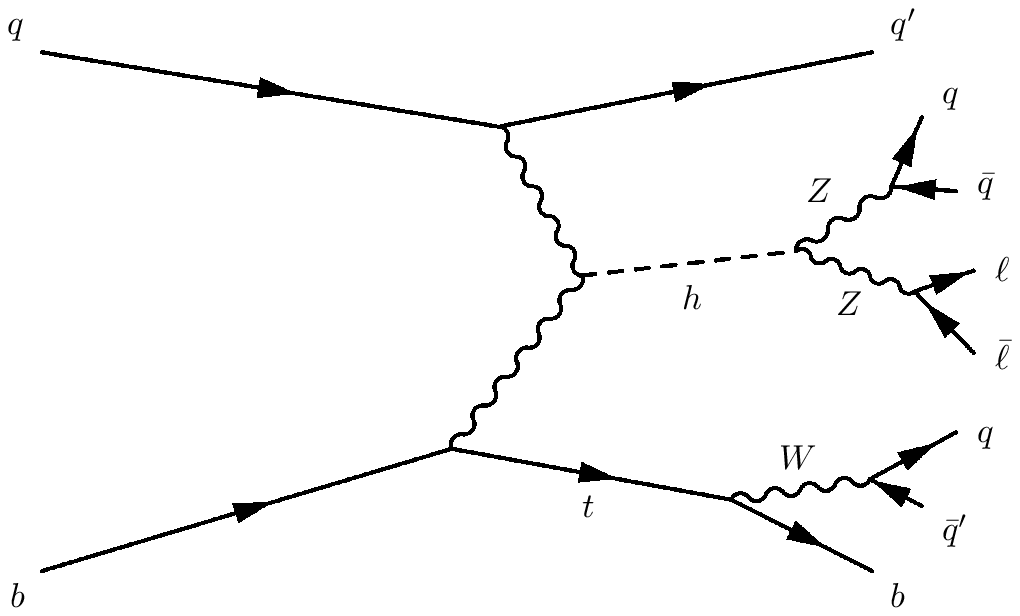}~~~
\includegraphics[angle=0,width=0.4\textwidth]{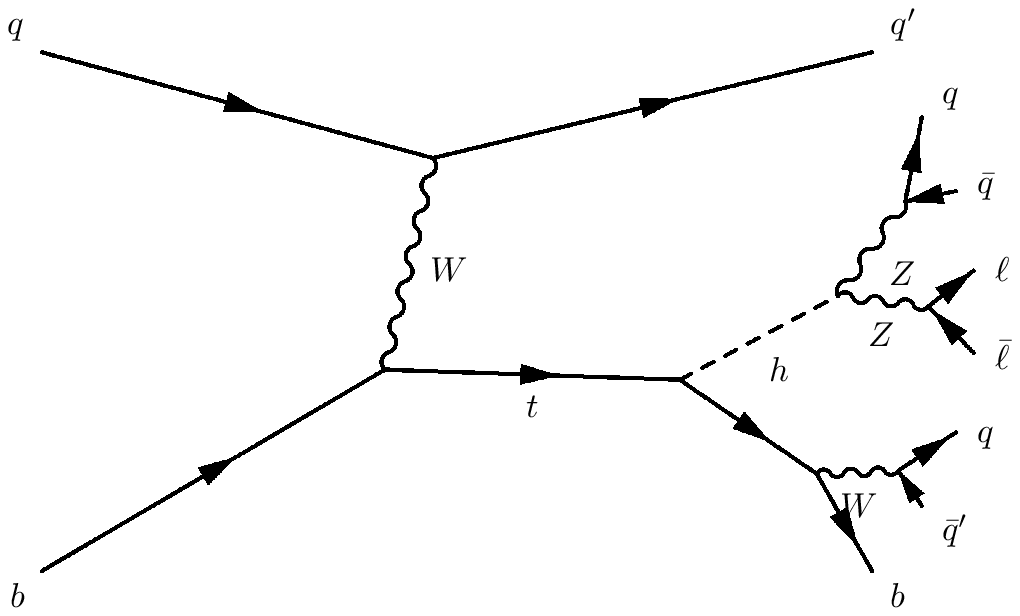}
\caption{Extended Feynman diagrams for the $tjh$ signal where the $h\to ZZ$ decay channel is shown.}
\label{fig:tjhFD}
\end{center}
\end{figure}

Next we reconstruct the top from a $b$ jet and two other light jets.  Finally, we once again tag a jet with high rapidity and high $p_{T}$.  For the highest rapidity jet, we impose a minimum rapidity of $2.5$.  The results of this series of cuts are given in Table~\ref{tab:htoZZ-tjh}.  It can be seen that this channel produces favorable results for extracting the Higgs signal when the Higgs has an appreciable branching fraction to $ZZ$.  This favors the $200$ GeV and $150$ GeV cases.  The $180$ GeV Higgs mass is above the $WW$ threshold and below the $ZZ$ threshold so the decay to WW dominates.  For even larger Higgs masses, the branching fraction to $ZZ$ asymptotically approaches roughtly $33\%$.  For $m_{h}=200$ GeV we obtain a statistical significance of $3.9\sigma$ with $600$ fb$^{-1}$ of integrated luminosity.  The diagram for this process that gives the best detection significance is shown in Fig~\ref{fig:tjhFD}.

\begin{table}[htpb]
\begin{center}
\begin{tabular}{|c|c|c|c|c|c|c|c|c|c|}
\hline
Signal Event Rate at $600$ fb$^{-1}$ & \multicolumn{3}{c|}{$m_{h}=150$ GeV} & \multicolumn{3}{c|}{$m_{h}=180$ GeV} & \multicolumn{3}{c|}{$m_{h}=200$ GeV} \\
for $6j+2\ell$: &  \multicolumn{3}{c|}{$84$ events} & \multicolumn{3}{c|}{$51$ events} & \multicolumn{3}{c|}{$216$ events} \\
\hline
Cut & $S$ & $B$ & $\frac{S}{\sqrt{S+B}}$ & $S$ & $B$ & $\frac{S}{\sqrt{S+B}}$ & $S$ & $B$ & $\frac{S}{\sqrt{S+B}}$ \\
\hline
tagging: $6$ j, $2\ell$, $\le 1$ b & $11$ & $1500$ & $0.3\pm0.1$ & $9$ & $1500$ & $0.2\pm0.1$ & $53$ & $1500$ & $1.3\pm0.2$ \\
$|M_{\ell\ell~\text{and}~jj}-M_{Z}|<10$ GeV & $7$ & $1400$ & $0.2\pm0.1$ & $7$ & $1400$ & $0.2\pm0.1$ & $53$ & $1400$ & $1.4\pm0.2$ \\
$|M_{bjj}-m_{t}|<20$ GeV & $6$ & $1000$ & $0.2\pm0.1$ & $4$ & $1000$ & $0.1\pm0.1$ & $45$ & $1000$ & $1.4\pm0.2$ \\
$|M_{2j2\ell}-m_{h}|<10$ GeV & $5$ & $28$ & $0.9\pm0.4$ & $4$ & $100$ & $0.4\pm0.2$ & $41$ & $180$ & $2.8\pm0.5$ \\
$|\eta_{j}|>2.5$ & $3$ & $6$ & $1.0\pm0.6$ & $2$ & $20$ & $0.3\pm0.2$ & $35$ & $46$ & $3.9\pm0.7$ \\
\hline
\end{tabular}
\caption{Table of the cuts used to extract the $tjh$ signal with $h\to ZZ^{(*)}$ decay for Higgs masses of $150$, $180$, and $200$ GeV.  The event rates include the appropriate branching fractions to obtain a final state of $6$ jets and $2$ leptons. The number of signal and background events that pass the cut are given as well as the resulting statistical significance.  For a Higgs mass of $200$ GeV a statistical significance of $3.9\sigma$ is obtained with an integrated luminosity of $600$ fb$^{-1}$.} 
\label{tab:htoZZ-tjh}
\end{center}
\end{table}

The $tWh$ signal with $H\to ZZ^{(*)}$ cannot be used as effectively.  The $tWh$ signal with $600$ fb$^{-1}$ of integrated luminosity gives an order of only $10$ events before acceptance and isolation cuts.  Moreover this signal is dominated by the $t\bar{t}Zj$ background.

\section{Constraints on Beyond the SM couplings}
\label{sec:beyondtheSM}

The two contributing diagrams to Higgs production with a single top are proportional to the top Yukawa and the $WWh$ coupling, respectively. Therefore, the interference between these two diagrams depends on the sign of the $WWh$ coupling which allows a unique test of the SM prediction for this sign~\cite{Tait:2000sh}.

To illustrate, we parameterize the top Yukawa and gauge boson couplings as
\begin{eqnarray}
y_{t}&=&c_{t}y_{t}^{SM},\\
g_{WWh}&=&c_w g_{WWh}^{SM},
\label{eq:c_t}
\end{eqnarray}

\noindent where $c_{t}>0$ and $c_{w}=\pm1$ such that $c_{t}=c_{w}=1$ for the SM~\footnote{The top quark Yukawa coupling must be positive since its sign is also that of the fermion mass which is fixed by vacuum stability.}.  It is expected that with the luminosity we adopt for this study, the magnitude of the $WWh$ coupling will be measured precisely in the $WW$ fusion process~\cite{Eboli:2000ze}.  However, there is no direct way of determining the sign of the $WWh$ coupling via the partial width of $h\to W^+ W^-$~\footnote{In principle, the loop induced decay of $h\to \gamma \gamma$ can provide this information~\cite{Ellis:1975ap}, but this decay is also sensitive to other states in the loop that may include new physics contributions, potentially masking the interference effect.}.  A sign change for the $WWh$ coupling can occur in multi-Higgs doublet models if the VEV direction in $\phi_1,\phi_2$ space is anti-aligned to the physical Higgs state~\cite{Barger:2009me}.  For unitarity, there must be a second Higgs boson whose coupling to the gauge bosons has a positive sign.  Therefore, if a negative sign of the gauge boson coupling to the Higgs is found, an additional CP even Higgs boson must exist.  

In the following exploration of the top Yukawa strength, we adopt an absolute $WWh$ coupling equal to that of the SM.  If a coupling departure from the SM is found, it will be a reduction if the model contains doublets and/or singlets.

\begin{figure}[htpb]
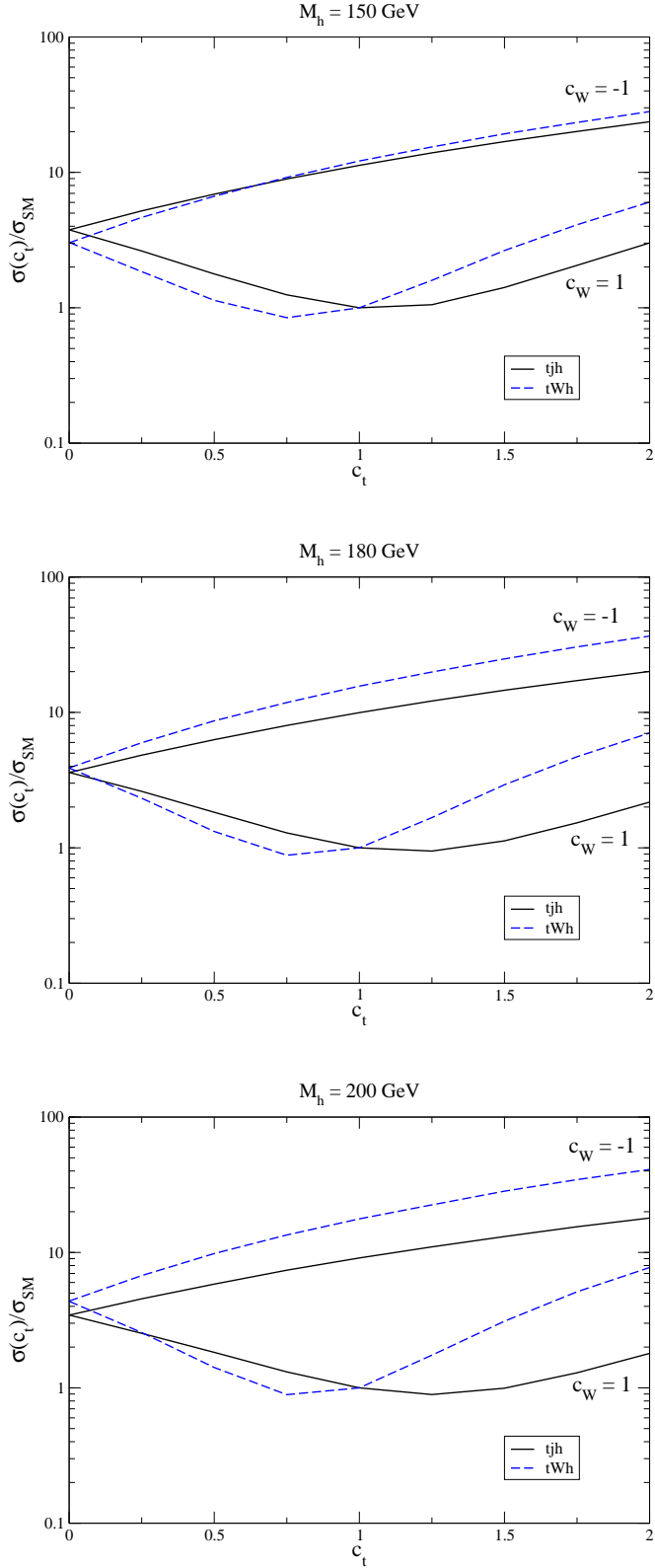

\begin{center}
\includegraphics[angle=0,width=0.53\textwidth]{figs/Yukawa-cross-rel-150.eps}\\
\vspace{0.7cm}
\includegraphics[angle=0,width=0.53\textwidth]{figs/Yukawa-cross-rel-180.eps}\\
\vspace{0.7cm}
\includegraphics[angle=0,width=0.53\textwidth]{figs/Yukawa-cross-rel-200.eps}
\caption{Cross sections of Higgs production methods with associated top quarks with respect to the ratio of the top Yukawa coupling to the SM Yukawa coupling, $c_{t}$.  The cross sections are scaled by that of the SM.  There is a significant enhancement in the cross section when the sign of the $WWh$ coupling is negative.}
\label{fig:crossYukawa-ratio}
\end{center}
\end{figure}

In Fig.~\ref{fig:crossYukawa-ratio} we show the effect of changing the top Yukawa scaling with both signs for the $WWh$ coupling on the various Higgs production mechanisms. A significant increase in the cross sections occurs when the sign of the $WWh$ coupling is negative.  For example, with the SM top Yukawa and $c_{w}=-1$ the cross sections for the $tjh$ and $tWh$ channels increases by a factor of approximately $10$.  Accordingly, the integrated luminosity needed for the measurement of $c_{t}$ could be substantially less.

Such a change in the top Yukawa and gauge boson coupling may also change the distributions of the final state particles which may potentially influence the optimal cuts.  To quantify this, we repeated the analyses in Section~\ref{sec:collstudy} with several values of $c_{t}$ (in intervals of $0.25$) and signs of $c_{w}$.  This allows us to find a range of $c_{t}$ that yield a $>3\sigma$ significance with $600$ fb$^{-1}$ of integrated luminosity.  Table~\ref{tab:ctscan} shows the ranges of $c_{t}$ without the $c_{w}$ sign change.  When the sign of $c_{w}$ is flipped we find a statistical significance above $3\sigma$ for both signals and all Higgs masses studied.  This shows that the Higgs production with an associated single top can provide a definitive test of both the top quark Yukawa coupling and the sign of the $WWh$ coupling.

\begin{table}[htpb]
\begin{center}
\begin{tabular}{|c|c|c|c|}
\hline
Signal & $M_{h}=150$ GeV & $M_{h}=180$ GeV & $M_{h}=200$ GeV \\
\hline
$tjh~(c_{w}=1)$ & $1.75<c_{t}$ & $2.75<c_{t}$ & All $c_{t}$ \\
$tjh~(c_{w}=-1)$ & All $c_{t}$ & All $c_{t}$ & All $c_{t}$ \\ 
\hline
$tWh~(c_{w}=1)$ & $c_{t}<0.25$, $1.50<c_{t}$ & $c_{t}<0.25$, $1.50<c_{t}$ & $1.75<c_{t}$ \\
$tWh~(c_{w}=-1)$ & All $c_{t}$ & All $c_{t}$ & $0.25<c_{t}$ \\
\hline
\end{tabular}
\caption{Ranges of $c_{t}$ above $3\sigma$ with $600$ fb$^{-1}$ of integrated luminosity using the same cuts in Section~\ref{sec:collstudy} for both single top signals.  The quoted boundaries are inclusive so it is possible to obtain $3\sigma$ significance with values of $c_{t}$ slightly beyond these ranges.  When the sign of the $WWh$ is negative, most of the values of $c_{t}$ yield a statistical significance above $3\sigma$ for both signals and all Higgs masses.}
\label{tab:ctscan}
\end{center}
\end{table}

\section{Conclusions}

In this paper we studied Higgs production with an associated single top quark for Higgs masses in the range of $120$ GeV to $200$ GeV.  After making acceptance cuts to extract the signals from backgrounds, our conclusions are:

\begin{itemize}
\item The signal of the $120$ GeV Higgs is overwhelmed by QCD backgrounds, consistent with the findings of Ref.~\cite{Maltoni:2001hu}.

\item Overall, the best topology for isolating the signature of associated single top and Higgs production is via the $qb\to tjh$ subprocess with a hadronically decaying top quark and $h\to ZZ^{(*)}\to \ell^+ \ell^- + jj$.  For a SM Higgs with $m_{h}=200$ GeV we found a significance of $3.9\sigma$ with $600$ fb$^{-1}$ of integrated luminosity.  

\item Further, we also found that these signals can be a definitive test for the overall sign for the $WWh$ coupling.  Specifically, if the $WWh$ coupling is opposite in sign to the SM and the top quark Yukawa is the same as the SM, we found up to an order of magnitude increase in the overall event rate of single top and Higgs production at the LHC.  This provides a statistical significance above $5\sigma$ for all of the Higgs masses studied.  
\item If the sign of the $WWh$ coupling is negative, then a heavier scalar state must exist with a positive $WWh$ coupling to unitarize the model.  This would provide evidence for the existence of an extended Higgs sector.

\item We have determined the ranges in which the top quark Yukawa can be probed to $3\sigma$ with $600$ fb$^{-1}$ of integrated luminosity assuming either sign of the $WWh$ coupling; see Table~\ref{tab:ctscan}.

\end{itemize}

Our study shows that single top plus Higgs production should be observable at the LHC with large integrated luminosity and provide important insightes about the Higgs sector.  More detailed studies including detector simulations are warranted.

\section{Acknowledgments}

We thank Qing-Hong Cao, Ian Low, Carlos Wagner and Ed Berger for helpful discussions.  This work was supported in part by the U.S. Department of Energy under grants No. DE-FG02-95ER40896, DE-FG02-05ER41361, DE-FG02-08ER41531, and Contract DE-AC02-06CH11357, by the Wisconsin Alumni Research Foundation, and by the National Science Foundation grant No. PHY-0503584.

\bibliographystyle{h-physrev}
\bibliography{singletop}
\newpage

\end{document}